\documentclass[twocolumn,superscriptaddress,prl]{revtex4}
\usepackage{amssymb}
\usepackage{graphicx}
\usepackage{dcolumn}
\usepackage{bm}
\usepackage{amsmath}

\def\bb{\mathbf{b}}
\def\bp{\mathbf{p}}

\def\Prob{\mathrm{Prob}}

\usepackage{wasysym}
\usepackage[normalem]{ulem}
\usepackage[usenames,dvipsnames]{color}
\begin{document}

\title{Asymptotic Cellular Growth Rate as the Effective Information Utilization Rate}

\author{R. Pugatch}
\affiliation{School of Natural Sciences, Simons Center for Systems Biology, Institute for Advanced Study, Princeton, NJ 08540, USA}
\author{N. Barkai}
\affiliation{Department of Molecular Genetics, Weizmann Institute of Science, Rehovot 76100, Israel}
\author{T. Tlusty}
\affiliation{School of Natural Sciences, Simons Center for Systems Biology, Institute for Advanced Study, Princeton, NJ 08540, USA}

\begin{abstract}
We study the average asymptotic growth rate of cells in randomly fluctuating environments, with multiple viable phenotypes per environment. We show that any information processing strategy has an asymptotic growth rate, which is the sum of: (i) the maximal growth rate at the worst possible distribution of environments, (ii) relative information between the actual distribution of environments to the worst one, and (iii) information utilization rate, which is the information rate of the sensory devices minus the ``information dissipation rate'', the amount of information not utilized by the cell for growth. In non-stationary environments, we find that the optimal phenotypic switching times equally partition the information dissipation rate between consecutive switching intervals.
\end{abstract}

\maketitle

In order to grow, cells need to respond to the environment by choosing their phenotype. When external conditions change, cells that respond on time with the appropriate phenotypic response grow faster than cells that are slow or erroneous in their response. The phenotypic response occurs in a broad range of time scales, from seconds in certain metabolic switchings, to minutes (e.g. in the lag phase), to hours. Cells typically subjugate their gene expression to the environment via elaborate signal transduction networks that convey and process external cues to form the appropriate response. For example, S. Cerevisiae ribosome production levels, which are estimated to consume more than $60\%$ of the available free energy \cite{RibosomeProduction} are determined by the environment in a feed-forward loop \cite{SagiNaama}. Evidently, more information can potentially facilitate faster growth, but it is unclear how to quantify its actual utilization, as the information can also be ignored or misused.

Shannon noticed that the problem of communication can be addressed on three levels \cite{Shannon}: ``$(i)$ how accurately can the symbols of communication be transmitted? (the technical problem); $(ii)$ how precisely do the transmitted symbols convey the desired meaning? (the semantic problem); $(iii)$ how effectively does the received meaning affect conduct in the desired way? (the effectiveness problem)''.
In this work, we will address this problem in the context of cell growth in fluctuating environments. We show that it is possible to quantify both the information rate (the technical problem), and how well this information is utilized to allow for growth (the effectiveness problem). When the switching strategy is optimal, all the information is utilized for increasing the asymptotic growth rate \cite{Kelly,Cover,Leibler,Lachman}, a bound recently improved in the case of correlated environments and phenotypic switching with memory \cite{Oliv}.

Our goal is to consider the generic case where the utilization rate is not necessarily optimal and there is more than one viable phenotype per environment.  We derive a relation between the asymptotic growth rate (AGR) of a population of cells growing in a randomly fluctuating environment \cite{Leibler}, the amount of information the cells acquire about their environment and the manner in which they utilize this information for growth. We focus on a simplified scenario which can serve as a non-trivial starting point for studying this question in experiments. Thus we assume spatial homogeneity (well-mixing) conditions, an enforced external environment, unaffected by the cell's response, and a large population of cells. Furthermore, we assume rapid response of the cells to significant changes in their external environment.

To obtain the desired relation, we solve two optimization problems that represent two extreme scenarios. In the first scenario, the cells choose their phenotype without any information on the environment. The optimality criterion in this case is the game-theoretic min-max solution, $\Lambda_{\text{game}}$, the minimal growth rate the cell can secure irrespective of the environment. It serves as a lower bound that any useful information utilization strategy has to outperform. The game theoretic strategy is optimal if the worst case scenario is realized, and is guaranteed to yield better performance otherwise. Thus, a strategy that employs information should be compared with the guaranteed baseline we get by playing the game-theoretic solution. If we get less, then the information is not only useless but even harmful. We stress that this lower bound on the performance of information utilization strategies comes from a design, i.e. evolution, constraint and does not imply that nature is actively trying to annihilate cells.

The second scenario is optimal utilization rate of the available information, where the cell employs all its available information, albeit noisy, for growth. The asymptotic growth rate (AGR) in this case, $\Lambda_{\text{opt}}$, is an upper bound for the performance of any actual utilization strategy \cite{remarkAboutMaxUtilizationAGR}. We use these two bounds to reach our main result. By solving for these two extreme strategies we are able to decompose the average AGR of any given information utilization strategy $\Lambda$ into a sum over entropy rates:
\begin{eqnarray}\label{eq1a}
\Lambda=D\left(\bp||\bp^*\right)+U+\Lambda_{\text{game}} \leq \Lambda_{\text{opt}},
\end{eqnarray}
where $\bp^*$ is the least favorable distribution of environments, $D(\bp||\bp^*)$ is the relative information between the actual distribution of environment $\bp$ and $\bp^*$, $0 \leq U \leq I$ is the information utilization rate which is tightly bounded by the information channel rate $I$, and $\Lambda_{\text{opt}}=D\left(\mathbf{p}||\mathbf{p}^*\right)+I+\Lambda_{\text{game}}$.

\emph{Formal setting of the asymptotic growth problem.--} Consider a set of distinct viable environments denoted by an index $E \in \{1,...,n\}$. We assume that the environments change stochastically and have a stationary probability distribution function $\bp$, with $p^{\text{e}}_i=\Prob(E=i)$. The set of available phenotypes is denoted by an index $B \in \{1,...,m\}$, $m \leq n$. $\bb$ is a probability vector with elements  $b_k=\Prob(B=k)$ --- the probability that a random cell drawn from the population will present phenotype $k$. The side-information $S \in \{1,...,n\}$ is the output of the information channel. The $n \times m$ growth rate matrix $M_{ik}>0$ is the growth rate of phenotype $k$ in environment $i$. We write the continuous population dynamics equation as
\begin{equation}
\frac{dN_k}{dt}=M_{i(t)k} N_k(t) + \sum_{l=1}^{m} \Omega^{j(t)}_{kl}N_l(t), \label{eq1}
\end{equation}
with $N_k(t)$ being the number of cells with phenotype $k$ at time $t$. Also, $i(t)$ is the (random) environment index at time $t$, and $\Omega^{j(t)}$ is an $m \times m$ phenotypic transition rate matrix which depends on the current state of knowledge of the cell about the environment, $S=j(t)$ \cite{remarkTransitionMatrix}.

The information channel in our model is an abstract representation of the molecular machinery in the cell that convey information about the external environment. This information can be the internal concentrations of external metabolites transported to the cell \cite{BayesInferenceBindingPNAS}, but it can also be a transduction network that relays a signal triggered by a binding event to a membranal protein sensor. The information rate through the channel can be evaluated from the joint probability distribution for $i$ and $j$ \cite{BayesInferenceBindingPNAS}. With the additional assumption that $\Omega^{j(t)}$ is irreducible, i.e. every phenotype is accessible from any other phenotype, a unique zero eigenvalue exists and the corresponding eigenvector is the steady state phenotype probability distribution $\bb^{j(t)}$, which is a vector whose $k$-th element is $b_{B=k|S=j}$ --- the probability for the cell to choose phenotype $B=k$ given the information that the environment is $S=j$.

\emph{From continuous to discrete dynamics via the strong mixing approximation.--} Assuming that the average mixing rates $K_j$ of each transition matrix $\Omega^{j}$ \cite{Kemeny}, are much larger than the maximal growth rate i.e. $K_j \gg \max_i M_{ii}$ for all $j$, we can approximate Eq. (\ref{eq1}) by its discrete version, where we replace each transition matrix $\Omega^{j(t)}$ with the matrix that projects the population to the corresponding steady-state, $\bb^{j(t)}$ of $\Omega^j$. Thus,
\begin{eqnarray}
\nonumber N^{'}_k(t+\Delta T)=N(t) b_{k|j(t)}, \\
N_k(t+\Delta T)=O_{i(t)k} N^{'}_k(t+\Delta T), \label{eq2}
\end{eqnarray}
where $N(t)=\sum_k N_k(t)>0$ is the total population at time $t$ and $O_{ik}=e^{M_{ik}\Delta T}$ is the multiplicative growth factor, which measures by how much phenotype $k$ changed in environment $i$. The first step of Eq. (\ref{eq2}) accounts for the effect of a sudden change in the information $j$, e.g., due to a change in the environment. The second step of Eq. (\ref{eq2}) computes the change in the number of cells per phenotype after a time $\Delta T$ in the new environment.

The fold change in the population size after a time $t$ is $F(t)=\log\frac{N(t)}{N(t=0)}$. The asymptotic growth rate (AGR) is defined by $\Lambda =  \lim_{t \rightarrow \infty} \frac{1}{t} F(t) = \lim_{t \rightarrow \infty} \frac{1}{t} \log \frac{N(t)}{N(0)}$ \cite{remarkAGR}. The relation between the fold change after a switching event, $\log\frac{N(t)}{N(t-1)}$, and the overall fold change is given by $F(t)=\log \frac{N(t)}{N(t-1)}+\log \frac{N(t-1)}{N(t-2)}+\ldots+\log \frac{N(1)}{N(0)}$. It follows that the time averaged AGR from $t=0$ to $t=T$ is
\begin{equation}\label{eq4}
\Lambda_{\left\lfloor T \right\rfloor} = \frac{1}{\left\lfloor T \right\rfloor}\sum_{t=1}^{\left\lfloor T \right\rfloor} \log \left(\sum_{k} O_{i(t)k}b_{k|j(t)} \right),
\end{equation}
where $\left\lfloor T \right\rfloor=\left\lfloor\frac{T}{\Delta T}\right\rfloor$ is the number of environment switching events up to time $T$. Assuming ergodic environment statistics and taking the limit $T \rightarrow \infty$, we can replace the time average by an ensemble average,
\begin{equation}\label{eq5}
\Lambda = \sum_{ij} p_{j|i}p^{\text{e}}_{i} \log \left(\sum_{k} O_{ik}b_{k|j} \right). 
\end{equation}
$p_{S=j|E=i}p^{\text{e}}_{i} = p_{E=i|S=j}p^{\text{s}}_{j} $ is the joint probability distribution for the environment $E=i$ and the side-information $S=j$, where $p_{E=i|S=j}$ is the probability for the environment to be $E=i$ given that the side information is $S=j$, $p_{S=j|E=i}$ is the probability for the side information to be $S=j$ given that the environment is $E=i$, and $p^{\text{s}}_j$ and $p^{\text{e}}_i$ are the marginal probabilities for the side information and for the environment, respectively. The values of $p_{E=i|S=j}$ for all $i$'s and $j$'s reflect the quality of the information channel; e.g., a zero-loss channel is given by $p_{E=i|S=j}=\delta_{ij}$. In the absence of side-information, Eq. (\ref{eq5}) reduces to $\Lambda(\bp,\bb) = \sum_{i} p^{\text{e}}_{i} \log \left(\sum_{k} O_{ik}b_{k}\right)$, where $b_k$ is the probability for phenotype $k$.  Eq. \ref{eq5}, the starting point of our analysis, was previously discussed in \cite{Lachman,Cover,Oliv}.

To relate the AGR and information utilization rate, we show that Eq. (\ref{eq5}) can be decomposed, under some restrictions on the matrix $O$, into a sum of entropy rates. Previously, this was done only for a class of diagonal $O$ matrices called ``fair'' (see \cite{Cover}, p. 163). However, the assumption that for any environment there is a single viable phenotype is too strict to be useful. Typically, there are numerous phenotypes capable of growing in a given environment at different growth rates.  Our goal then is to present a generalization applicable for generic non-diagonal matrices. For this purpose, we first introduce two different optimality criteria which serve as upper and lower bound for the performance of any information utilization strategy.

\emph{The optimal solution.--} To find the optimal information utilization strategy given the channel performance and the environment distribution, we equate to zero the derivative of the average asymptotic growth rate in Eq. (\ref{eq5}) with respect to the information utilization strategy $b_{B=k|S=j}$, employing Lagrange multipliers to keep $b_{B=k|S=j}$ within the probability simplex. The optimal utilization strategy derived in \cite{suppmat}, is $b^{\text{opt}}_{B=k|S=j}=\sum_l W_{kl}p_{E=l|S=j}$, where $W_{kl}\equiv \frac{O^{-1}_{kl}}{\sum_m O^{-1}_{ml}}$. We require that ${\sum_m O^{-1}_{ml}}$ is strictly positive for all $l$'s, which implies that the matrix $W^{-1}$ is stochastic \cite{suppmat}. The meaning of this requirement will become evident in what follows. Note that the matrix $W$ itself is not necessarily stochastic as it typically contains negative elements \cite{remarkBirkhoff}. We define the positive diagonal matrix $d$ by $d_{ij}=\delta_{ij}(\sum_k (O^{-1})_{ki})^{-1}$. Then, $W=O^{-1}d$ \cite{RelevantLim}.

\emph{Min-max solution--.} Next, we introduce the game-theoretic optimal solution also known as the min-max point or Nash equilibrium \cite{GameTheoryBook}. Consider the AGR as a payoff function in a zero-sum game against nature \cite{GameAgainstNature}. For the technical purpose of finding the min-max point, we pretend that nature is an adversary that is trying to minimize the average cellular AGR, $\Lambda(\bb,\bp)$, while the cell, in the absence of any information, tries to maximize $\Lambda$. Both probability distributions $\bb$ and $\bp$ are independently determined. From the previous assumption that $W^{-1}$ is a strictly positive stochastic matrix it follows that the matrix $O$ has a single Nash equilibrium \cite{suppmat}, which is totaly mixing in the phenotypic strategy space \cite{Karlin,PositiveMatricesBook}, i.e. there are no phenotypes that should be avoided under all circumstances (``bad phenotypes''). Note that the set of matrix games with a unique Nash equilibrium is dense in the set of all matrix games \cite{Karlin} (p. 76).

By starting from a rectangular matrix $O$ and eliminating all the convexly dominated phenotypic strategies, one can obtain a reduced square sub-matrix $\tilde{O}$ of size $r \times r$, $r \leq m$, called the \textit{essential part of the game} \cite{Karlin}. This elimination process is always possible by finding the min-max vectors of $O$, and then erasing all rows and columns in $O$ not used by the min-max strategy vectors \cite{Karlin,suppmat}.  From hereon we will assume that $O$ is already essential \cite{LeftInverseRemark}. For an algorithm to perform the reduction applicable for any growth matrix see \cite{suppmat}, section (iv).

The min-max (Nash) solution is characterized by a pair of probability distributions $\bp^*=e^{\Lambda_{\text{game}}}\mathbf{1}^t O^{-1}$ for nature, and $\bb^*=e^{\Lambda_{\text{game}}}O^{-1}\mathbf{1}$ for the phenotypes, where $\Lambda_{\text{game}}=-\log \left(\mathbf{1}^t O^{-1} \mathbf{1} \right)$ is the \textit{game value}, i.e. the minimal AGR that the cell can secure, by playing Nash, irrespective of nature's strategy \cite{GameTheoryBook}, and $\mathbf{1}$ is the all-ones column vector. In other words, we find that $\Lambda(\bb,\bp^*) \leq \Lambda(\bb^*,\bp^*)$ and $\Lambda(\bb^*,\bp) \geq \Lambda(\bb^*,\bp^*)$ with $\Lambda_{\text{game}} \equiv \Lambda(\bb^*,\bp^*)$. The game-theoretic, zero-information optimal solution serves as a minimal benchmark for all other utilization strategies that do rely on information, since it guarantees a minimal level of AGR which can only improve if the cells stick to this strategy while nature deviates from the worst case scenario.

\emph{General solution--.} Upon introducing the optimal utilization strategy for a given information channel with its associated optimal AGR, $\Lambda_{\text{opt}}$, and the game-theoretic optimum with its associated AGR, $\Lambda_{\text{game}}$, we now proceed to derive our main result. Our starting point is Eq. (\ref{eq5}). Let us define the channel matrix $\Pi$ such that $\Pi_{ij}=p_{E=i|S=j}$ and the utilization matrix $B$ such that $B_{kj}=b_{B=k|S=j}$, where the $j$-th column of $B$ is $\bb_{\cdot|S=j}$. Rewriting Eq. (\ref{eq5}) in matrix form, the optimal utilization strategy for a given channel matrix $\Pi$ satisfies $B^{\text{opt}}=W\Pi=O^{-1}d\Pi$, where $(B^{\text{opt}})_{kj} \equiv b^{\text{opt}}_{B=k|S=j}$. We invert this relation to find $\Pi=W^{-1}B^{\text{opt}}=d^{-1}OB^{\text{opt}}$. Replacing $B^{\text{opt}}$ with an arbitrary utilization strategy matrix $B$ we obtain that $\Pi'=W^{-1}B=d^{-1}OB$. Thus, $\Pi'$ is the channel performance matrix for which a cell with an arbitrary utilization strategy $B$ is optimal \cite{remarkDpositive}. We can now rewrite Eq. (\ref{eq5}) as follows:
\begin{eqnarray}\label{eq7}
\nonumber \Lambda&=&\sum_{ij} p_{i|j} p^{\text{s}}_j\log(\sum_{k} d_{ii} d^{-1}_{ii} O_{ik}b_{k|j}) \\
&=&\sum_{ij} p_{i|j} p^{\text{s}}_j\log(d_{ii} p'_{i|j}).
\end{eqnarray}
Noticing that the game-theoretic optimal solution for nature is $p_i^*=\frac{1}{\sum_k d^{-1}_{kk}}d^{-1}_{ii}$, we can replace $d_{ii} p'_{i|j}$ in the logarithm by a product of four terms: $\Lambda = \sum_{ij} p_{i|j} p^{\text{s}}_j\log\left( \frac{p^{\text{e}}_i}{p^*_i}\frac{p'_{i|j}}{p_{i|j}} \frac{p_{i|j}}{p^{\text{e}}_i} \frac{1}{\sum_i d_{ii}^{-1}} \right)$. By summing separately each term in the logarithm of the product and using the relation $p^{\text{e}}_i p_{S=j|E=i}=p^{\text{s}}_j p_{E=i|S=j}$, we obtain our main result:
\begin{eqnarray}\label{eq11}
\nonumber \Lambda&=&D\left(\bp||\bp^*\right)+I\left(E;B\right) \\ &-&\sum_j p^{\text{s}}_j D\left(\bp_{\cdot|j}||\bp'_{\cdot|j}\right)+\Lambda_{\text{game}}.
\end{eqnarray}
The term $D\left(\bp||\mathbf{p^*}\right)=\sum_i p^{\text{e}}_i \log\frac{p^{\text{e}}_i}{p^*_i}$ is the relative information (Kullback-Leibler divergence) between nature's actual distribution of environments $\bp$ and the worst possible one $\bp^*$. It measures how much the AGR can potentially gain when nature is not playing the worst possible strategy. The term $-\sum_j p^{\text{s}}_j D\left(\bp_{\cdot|j}||\bp'_{\cdot|j}\right) \leq 0$ is the \textit{information dissipation rate}, the amount of information not utilized for growth. It vanishes if the utilization strategy optimally fits the channel performance. $I(E;B)=D\left(p_{E=i|S=j}p^{\text{s}}_j||p^{\text{e}}_i p^{\text{s}}_j \right)$ is the mutual information that measures how much information is conveyed by the channel.

In terms of Shannon's hierarchy, $I$ measures how accurately the different environment states are transmitted, while the third (penalty) term in Eq. (\ref{eq11}) measures how effectively the received information affects the phenotypic switching to allow for faster growth. The combined term
\begin{eqnarray}\label{eq12b}
U(E;B)=I(E;B)-\sum_j p^{\text{s}}_j D(\bp_{\cdot|j}||W^{-1}\bb_{\cdot|j}),
\end{eqnarray}
that appears in Eq. (\ref{eq11}) can be interpreted as the \emph{effective information utilization rate} which is the sum of the channel information rate minus the information dissipation rate \cite{DwellTimeFixUnits}.

It is instructive to consider the case where no side information is available (bet-hedging). In this case, Eq. (\ref{eq7}) yields the average AGR:
\begin{eqnarray}\label{eq12}
\Lambda_{\text{bet}}=D\left(\bp||\bp^*\right)-D\left(\bp||\bp'\right)+\Lambda_{\text{game}},
\end{eqnarray}
where $\bp'=S^{-1}\bb$ is the distribution of environments for which the bet-hedging strategy $\bb$ is optimal. The difference between the average AGR with and without an information channel is given by $\Delta \Lambda_{\text{ch}}=\Lambda-\Lambda_{\text{bet}}=-\sum_j p^{\text{e}}_j D\left(\bp_{\cdot|S=j}||\bp'_{\cdot|S=j}\right)+D\left(\bp||\bp'\right)+I(E;B)$. Comparing between optimal strategies, the first and the second terms vanish and we obtain that $\Delta \Lambda^{\text{opt}}_{ch} = I(E;B)$, i.e. the channel capacity. This is in accord with previously obtained tight upper bounds presented in \cite{Kelly,Leibler} and in particular in \cite{Lachman,Cover} for non diagonal growth matrices.

\emph{Interpretation in terms of correlated equilibrium--.} Correlated equilibrium (CE) is a generalization of Nash-equilibrium that allows for correlations between players \cite{AumanCorrEq}. Consider two players (in our case, nature and the cell) that choose their strategies using marginal probability distributions derived from a joint distribution $J$, $b_k=\sum_l J_{kl}$ and $p_l=\sum_k J_{kl}$. The players are in correlated equilibrium (CE) if, assuming one player fixes his strategy to the marginal of $J$, the other player has no incentive to deviate from its marginal either. It follows that a Nash-equilibrium is also a CE with $J_{kl}=b^*_k p^*_l$. When cells base their phenotypic switching on measurements, albeit noisy, of the external environments, they will always grow faster, as long as they respond at a rate that is faster or comparable to the switching rate of the environment (as assumed throughout our analysis). This suggests the information utilization strategy as a natural pathway to CE \cite{corrEqRemark}.

\emph{Metabolic pathways with binary alternatives--.} Consider a population of Ecoli bacteria in a minimal M9 medium (environment $E=1$). The second environment, $E=2$, is the also an M9 medium supplemented with an additional essential amino-acid say Histidine ($His$). It follows that the second medium has two alternative nitrogen sources (NH$_4$Cl from the M9 medium, and $His$). The net rate of accumulating $His$ in a bacterium residing in $E=2$ is the sum of the rate of uptake and the rate of de-novo synthesis. The dependence of the growth rate on the balance between the two alternatives, uptake and production, is not necessarily linear due to the overhead cost of the production channel. Let the maximally growing phenotype in $E=2$ be a combination $V_{\text{His}}^{\text{opt}}=\alpha^{\text{opt}} V_{\text{uptake}}+(1-\alpha^{\text{opt}}) V_{\text{synth}}$, with $0< \alpha^{\text{opt}} \leq 1$. We can now reduce the strategy space into two convexly dominant phenotypes: (i) cells with phenotype $B=1$ do not uptake the amino-acid but rather produce it. (ii) cells with phenotype $B=2$ regulate the synthesis of the amino-acid and the uptake to the optimal level $\alpha^{opt}$; The growth matrix is a $2 \times 2$ matrix with elements $O_{E=1,B=1}=e^{\nu \Delta T}$, $O_{E=1,B=2}=e^{-\mu \Delta T}$ (which represents dilution of the essential amino-acid that eventually leads to cell death), $O_{E=2,B=1}=e^{\nu \Delta T}$ and $O_{E=2,B=2}=e^{\mu \Delta T}$, where $\mu$ is the maximal growth rate in $E=2$ and $\nu<\mu$ is the maximal growth rate in $E=1$.
We suggest the following experiment to test our prediction \cite{suppmat}. Consider bacteria growing in a randomly generated sequence of media $E=1$ or $E=2$, with a user defined probability $\mathbf{p}$. The side-information and the phenotype should be measured throughout the population. Then each term in Eq. (\ref{eq11}) can be independently measured. If correct, a linear relation between the measured AGR, $\Lambda$, and the independently measured sum $D\left(\bp||\bp^*\right)+U$ is expected when we vary $\mathbf{p}$. The phenotype can be measured by monitoring the de-novo synthesis of $His$ in the two environments. The side information can be measured by the internal level of the transported $His$, e.g., by labeling the external $His$. The AGR can be independently measured from the population size \cite{suppmat}. For a a Monte-Carlo simulation of this experiment, a generalized scheme involving more than a binary choice, and for two other suggested examples for experiments see section (ii) and (v) in \cite{suppmat}.

\emph{Non-stationary environments--.} When the environment distribution varies at a rate lower than the phenotypic switching rates, $\Lambda(\bb(t),\bp(t))$ is still meaningful. The instantaneous loss function is then $L(t)=\Lambda_{\text{opt}}-\Lambda = \sum_{ij} p_{E=i|S=j}(t) p^{\text{s}}_j\log \frac{p_{E=i|S=j}(t)}{\sum_k W^{-1}_{ik}b_{B=k|S=j}(t)}$. The optimal strategy is to instantaneously tune the switching strategy such that it will track the environment given the channel performance. However, this is impractical as it takes time to sense the changes and respond. More realistically, the phenotype distribution may change only once in each time interval $[t_{\nu},t_{\nu+1}]$. To find the optimal switching strategy given this constraint, we minimize $\int_{t_{\nu}}^{t_{\nu+1}} L(t) dt - \lambda ({\sum_k b_{k|j} -1)}$, where $\lambda$ is a Lagrange multiplier. The resulting optimal strategy is $\sum_k W^{-1}_{ik} b_{B=k|S=j}=\frac{1}{t_{\nu+1}-t_{\nu}} \int_{t_{\nu}}^{t_{\nu+1}} dt p_{E=i|S=j}(t) \equiv \bar{p}_{E=i|S=j}$ i.e. a time average over the optimal instantaneous response. We can also ask what is the optimal switching times for a given phenotypic response, not necessarily optimal. The result, derived in \cite{suppmat} is to switch in a way that distributes the information dissipation rate evenly between consecutive switching events. It is possible to define a Riemannian metric over the strategy space, such that the infinitesimal distance squared between two strategies is given by $(\Delta L)^2=\sum_{ij} p^s_j \frac{\Delta p^{2}_{E=i|S=j}}{2 p_{E=i|S=j}}$ \cite{Akin}. Using this metric we can give a geometrical interpretation to the optimal switching times, namely that they are equidistant. In the limit of continuous optimal switching, the resulting trajectory in probability space is a geodesic line.

To conclude, we study the asymptotic growth rate of cells in fluctuating environments and find it can be decomposed into a sum of (i) the baseline growth rate at the worst possible condition (game value) $\Lambda_{\text{game}}$, (ii) the relative information between the actual environment distribution and the worst possible distribution $D(\bp||\bp^*)$, and (iii) the information utilization rate,
$U=I-\sum_j p^{\text{s}}_j D(\bp(\cdot|S=j)||\bp'(\cdot|S=j))$. $U$ is the difference between the information rate of the sensing device and the information dissipation rate. 
If the organism optimally utilizes the information from the sensors, then any bit of information gained can be directly translated to gain in the AGR.

Recently, several generalizations to the problem of population growth in varying environments have been examined \cite{Oliv}. In particular, it was shown, in the case of less-than-fatal penalty of non-optimal strategies, that the additional growth rate gained by adjusting the phenotypic response is bounded by the channel capacity $I$. Also discussed in \cite{Oliv} is modulation of the phenotype that also takes into account a feedback from past phenotypes. It was shown that the mutual information bound in the presence of this feedback can be further tightened by the directed information bound. Interestingly, the channel capacity bound $I$ we obtained can be restored if feedback in the phenotypic switching mechanism is avoided \cite{Massey}. The growth benefit of feedback exists when different environments are correlated in time. When the correlation decay time is smaller than the feedback delay, feedback becomes detrimental and better growth rates are attained without it. This might be the reason why transporters and ribosome production in several single-cell organisms are modulated by signal transduction networks that are feed-forward and lack global feedback \cite{transporterYeastPNAS,SagiNaama}.

Finally, when the environment is non-stationary, we found that the optimal strategy is to use the time-average of the instantaneous optimal strategy and to switch at times that distribute the loss evenly among the switching intervals. While we did not discuss how a cell can actually approach this optimal phenotypic response, recent studies suggest how a crucial quantity in this natural computation, namely the the conditional probability $p_{i|j}$, might be calculated by a network of interacting proteins \cite{BayesInferenceBindingPNAS,Kobayashi}.

We thank BingKan Xue for helpfull comments.


\begin{thebibliography}{29}
\expandafter\ifx\csname natexlab\endcsname\relax\def\natexlab#1{#1}\fi
\expandafter\ifx\csname bibnamefont\endcsname\relax
  \def\bibnamefont#1{#1}\fi
\expandafter\ifx\csname bibfnamefont\endcsname\relax
  \def\bibfnamefont#1{#1}\fi
\expandafter\ifx\csname citenamefont\endcsname\relax
  \def\citenamefont#1{#1}\fi
\expandafter\ifx\csname url\endcsname\relax
  \def\url#1{\texttt{#1}}\fi
\expandafter\ifx\csname urlprefix\endcsname\relax\def\urlprefix{URL }\fi
\providecommand{\bibinfo}[2]{#2}
\providecommand{\eprint}[2][]{\url{#2}}

\bibitem[{\citenamefont{Warner}(1999)}]{RibosomeProduction}
\bibinfo{author}{\bibfnamefont{J.}~\bibnamefont{Warner}},
  \bibinfo{journal}{Trends Biochem. Sci.} \textbf{\bibinfo{volume}{24}},
  \bibinfo{pages}{437} (\bibinfo{year}{1999}).

\bibitem[{\citenamefont{Levy and Barkai}(2009)}]{SagiNaama}
\bibinfo{author}{\bibfnamefont{S.}~\bibnamefont{Levy}} \bibnamefont{and}
  \bibinfo{author}{\bibfnamefont{N.}~\bibnamefont{Barkai}},
  \bibinfo{journal}{FEBS Letters} \textbf{\bibinfo{volume}{583}},
  \bibinfo{pages}{3974} (\bibinfo{year}{2009}).

\bibitem[{\citenamefont{Shannon and Weaver}(1949)}]{Shannon}
\bibinfo{author}{\bibfnamefont{C.~E.} \bibnamefont{Shannon}} \bibnamefont{and}
  \bibinfo{author}{\bibfnamefont{W.}~\bibnamefont{Weaver}},
  \emph{\bibinfo{title}{The Mathematical Theory of Communication}}
  (\bibinfo{publisher}{The University of Illinois Press, Urbana, Illinois},
  \bibinfo{year}{1949}).

\bibitem[{\citenamefont{Kelly}(1956)}]{Kelly}
\bibinfo{author}{\bibfnamefont{J.~L.} \bibnamefont{Kelly}},
  \bibinfo{journal}{Bell System Technical Journal}
  \textbf{\bibinfo{volume}{35}}, \bibinfo{pages}{917} (\bibinfo{year}{1956}).

\bibitem[{\citenamefont{Cover and Thomas}(2006)}]{Cover}
\bibinfo{author}{\bibfnamefont{T.~M.} \bibnamefont{Cover}} \bibnamefont{and}
  \bibinfo{author}{\bibfnamefont{J.~A.} \bibnamefont{Thomas}},
  \emph{\bibinfo{title}{Elements of Information Theory}}
  (\bibinfo{publisher}{Wiley Interscience}, \bibinfo{year}{2006}).

\bibitem[{\citenamefont{Kussell and Leibler}(2005)}]{Leibler}
\bibinfo{author}{\bibfnamefont{E.}~\bibnamefont{Kussell}} \bibnamefont{and}
  \bibinfo{author}{\bibfnamefont{S.}~\bibnamefont{Leibler}},
  \bibinfo{journal}{Science} \textbf{\bibinfo{volume}{309}},
  \bibinfo{pages}{2075} (\bibinfo{year}{2005}).

\bibitem[{\citenamefont{Donaldson-Matasci
  et~al.}(2009)\citenamefont{Donaldson-Matasci, Bergstrom, and
  Lachmann}}]{Lachman}
\bibinfo{author}{\bibfnamefont{M.~C.} \bibnamefont{Donaldson-Matasci}},
  \bibinfo{author}{\bibfnamefont{C.~T.} \bibnamefont{Bergstrom}},
  \bibnamefont{and} \bibinfo{author}{\bibfnamefont{M.}~\bibnamefont{Lachmann}},
  \bibinfo{journal}{Oikos} \textbf{\bibinfo{volume}{119}}, \bibinfo{pages}{219}
  (\bibinfo{year}{2009}).

\bibitem[{rem({\natexlab{a}})}]{remarkAboutMaxUtilizationAGR}
\bibinfo{note}{Note that $\Lambda_{\text{opt}}$ depends on the information
  channel and is peaked for a lossless information channel.}

\bibitem[{rem({\natexlab{b}})}]{remarkTransitionMatrix}
\bibinfo{note}{For all $j$ and $k$ the transition matrix satisfies that
  $\sum_{l} \Omega^{j(t)}_{lk} = 0$.}

\bibitem[{\citenamefont{Libby et~al.}(2007)\citenamefont{Libby, Perkins, and
  Swain}}]{BayesInferenceBindingPNAS}
\bibinfo{author}{\bibfnamefont{E.}~\bibnamefont{Libby}},
  \bibinfo{author}{\bibfnamefont{T.~J.} \bibnamefont{Perkins}},
  \bibnamefont{and} \bibinfo{author}{\bibfnamefont{P.~S.} \bibnamefont{Swain}},
  \bibinfo{journal}{Proc. Natl. Acad. Sci. USA} \textbf{\bibinfo{volume}{104}},
  \bibinfo{pages}{7151} (\bibinfo{year}{2007}).

\bibitem[{\citenamefont{Catral et~al.}(2010)\citenamefont{Catral, Kirkland,
  Neumann, and Sze}}]{Kemeny}
\bibinfo{author}{\bibfnamefont{M.}~\bibnamefont{Catral}},
  \bibinfo{author}{\bibfnamefont{S.~J.} \bibnamefont{Kirkland}},
  \bibinfo{author}{\bibfnamefont{M.}~\bibnamefont{Neumann}}, \bibnamefont{and}
  \bibinfo{author}{\bibfnamefont{N.~S.} \bibnamefont{Sze}},
  \bibinfo{journal}{Journal of Scientific Computing}
  \textbf{\bibinfo{volume}{45}}, \bibinfo{pages}{151} (\bibinfo{year}{2010}).

\bibitem[{rem({\natexlab{c}})}]{remarkAGR}
\bibinfo{note}{In a fixed environment with an exponential growth rate $\mu$,
  the asymptotic growth rate equals to $\mu$.}

\bibitem[{sup()}]{suppmat}
\bibinfo{note}{See supplementary material for derivation and further details.}

\bibitem[{rem({\natexlab{d}})}]{remarkBirkhoff}
\bibinfo{note}{In fact, if both $W$ and $W^{-1}$ are stochastic then it follows
  from the von-Neumann-Birkhoff theorem that $W$ is a permutation matrix, which
  implies that $O$ is diagonal up to permutation [19]}.

\bibitem[{Rel()}]{RelevantLim}
\bibinfo{note}{Another relevant limit of Eq. (6) is obtained when the
  information about the external environment is statistical (i.e. only the
  environment distribution is known). In this bet-hedging limit, we obtain that
  the optimal strategy is $\mathbf{b}^{\text{opt}}=W\mathbf{p}$, where
  $\mathbf{p}$ is the probability vector for the different environments [11].
  If $O$ is diagonal then $W$ becomes the identity matrix and we recover the
  proportional betting result of Kelly: $b^{\text{opt}}=p$ [4,5]}.

\bibitem[{\citenamefont{Zamir et~al.}(2008)\citenamefont{Zamir, Meshler, and
  Solan}}]{GameTheoryBook}
\bibinfo{author}{\bibfnamefont{S.}~\bibnamefont{Zamir}},
  \bibinfo{author}{\bibfnamefont{M.}~\bibnamefont{Meshler}}, \bibnamefont{and}
  \bibinfo{author}{\bibfnamefont{E.}~\bibnamefont{Solan}},
  \emph{\bibinfo{title}{Game Theory}} (\bibinfo{publisher}{Magnes, Hebrew
  University Press}, \bibinfo{year}{2008}).

\bibitem[{\citenamefont{Milnor}(1951)}]{GameAgainstNature}
\bibinfo{author}{\bibfnamefont{J.}~\bibnamefont{Milnor}}, \bibinfo{type}{Tech.
  Rep.} \bibinfo{number}{RM-679}, \bibinfo{institution}{RAND}
  (\bibinfo{year}{1951}).

\bibitem[{\citenamefont{Karlin}(1992)}]{Karlin}
\bibinfo{author}{\bibfnamefont{S.}~\bibnamefont{Karlin}},
  \emph{\bibinfo{title}{Mathematical Methods and Theory in Games, Programming,
  and Economics}} (\bibinfo{publisher}{Dover}, \bibinfo{year}{1992}).

\bibitem[{\citenamefont{Bapat and Raghavan}(1997)}]{PositiveMatricesBook}
\bibinfo{author}{\bibfnamefont{R.~B.} \bibnamefont{Bapat}} \bibnamefont{and}
  \bibinfo{author}{\bibfnamefont{T.~E.~S.} \bibnamefont{Raghavan}},
  \emph{\bibinfo{title}{nonnegative matrices and applications}}
  (\bibinfo{publisher}{Cambridge University Press}, \bibinfo{year}{1997}).

\bibitem[{Lef()}]{LeftInverseRemark}
\bibinfo{note}{Alternativly, if we choose not to eliminate ``bad phenotypes''
  that are always dominated irrespective of the environment, we can still
  proceed with the derivation interpreting $O^{-1}$ as the
  \emph{left-inverse}.}

\bibitem[{rem({\natexlab{e}})}]{remarkDpositive}
\bibinfo{note}{As shown in the supplementary material [11], the positivity of
  the diagonal of $d$ follows from the assumption that $O$ is totally mixing.
  This in turn makes $W^{-1}$ stochastic, and thus our interpretation of
  $W^{-1} \textbf{b}$ follows.}

\bibitem[{Dwe()}]{DwellTimeFixUnits}
\bibinfo{note}{To fix the units to rate we should divide by the dwell-time
  $\Delta T$}.

\bibitem[{\citenamefont{Aumann}(1974)}]{AumanCorrEq}
\bibinfo{author}{\bibfnamefont{R.}~\bibnamefont{Aumann}}, \bibinfo{journal}{J.
  Math. Econ.} \textbf{\bibinfo{volume}{1}}, \bibinfo{pages}{67}
  (\bibinfo{year}{1974}).

\bibitem[{cor()}]{corrEqRemark}
\bibinfo{note}{A simple extension of CE more suited to our purpose is
  constrained correlated equilibrium (CCE), which is a correlated equilibrium
  with imposed constraints on at least one of the player's strategies. Assuming
  the constraint strategy space of each player is smooth and compact, a CCE is
  just a CE on the constraint strategy spaces}.

\bibitem[{\citenamefont{Akin}(1979)}]{Akin}
\bibinfo{author}{\bibfnamefont{E.}~\bibnamefont{Akin}},
  \emph{\bibinfo{title}{The Geometry of Population Genetics}}
  (\bibinfo{publisher}{Springe Verlag}, \bibinfo{year}{1979}).

\bibitem[{\citenamefont{Rivoire and Leibler}(2011)}]{Oliv}
\bibinfo{author}{\bibfnamefont{O.}~\bibnamefont{Rivoire}} \bibnamefont{and}
  \bibinfo{author}{\bibfnamefont{S.}~\bibnamefont{Leibler}},
  \bibinfo{journal}{J. Stat. Phys.} \textbf{\bibinfo{volume}{142}},
  \bibinfo{pages}{1124} (\bibinfo{year}{2011}).

\bibitem[{\citenamefont{Massey and Massey}(2005)}]{Massey}
\bibinfo{author}{\bibfnamefont{J.~L.} \bibnamefont{Massey}} \bibnamefont{and}
  \bibinfo{author}{\bibfnamefont{P.}~\bibnamefont{Massey}}, in
  \emph{\bibinfo{booktitle}{International Symposium on Information Theory}}
  (\bibinfo{year}{2005}), p. \bibinfo{pages}{157}.

\bibitem[{\citenamefont{Kuttykrishnana
  et~al.}(2010)\citenamefont{Kuttykrishnana, Sabinaa, Langtona, Johnstona, and
  Brenta}}]{transporterYeastPNAS}
\bibinfo{author}{\bibfnamefont{S.}~\bibnamefont{Kuttykrishnana}},
  \bibinfo{author}{\bibfnamefont{J.}~\bibnamefont{Sabinaa}},
  \bibinfo{author}{\bibfnamefont{L.~L.} \bibnamefont{Langtona}},
  \bibinfo{author}{\bibfnamefont{M.}~\bibnamefont{Johnstona}},
  \bibnamefont{and} \bibinfo{author}{\bibfnamefont{M.~R.}
  \bibnamefont{Brenta}}, \bibinfo{journal}{Proc. Natl. Acad. Sci. USA}
  \textbf{\bibinfo{volume}{107}}, \bibinfo{pages}{16743}
  (\bibinfo{year}{2010}).

\bibitem[{\citenamefont{Kobayashi}(2010)}]{Kobayashi}
\bibinfo{author}{\bibfnamefont{T.~J.} \bibnamefont{Kobayashi}},
  \bibinfo{journal}{Phys. Rev. Lett.} \textbf{\bibinfo{volume}{104}},
  \bibinfo{pages}{228104} (\bibinfo{year}{2010}).

\end{thebibliography}
\end{document}